\title{\bf Generic autonomous system approach to interacting dark energy models}
\author{Parth Shah$^{1}$, Gauranga C. Samanta$^{2}$, Kazuharu Bamba$^{3}$, R. Myrzakulov$^{4}$ \\
$^{1}$Department of Mathematics, VIT AP University,\\
Amaravati 522237, India \\
$^{2}$P.G. Department of Mathematics, Fakir Mohan University,\\
Balasore, Odisha 756019, India \\
$^{3}$Division of Human Support System, Faculty of Symbiotic Systems Science,\\
Fukushima University, Fukushima 960-1296, Japan \\
$^{4}$Eurasian National University,\\
Nur-Sultan 010008, Kazakhstan}

\documentclass[11pt,a4paper]{article}
\usepackage{amsmath, amsthm, amsfonts}
\usepackage[margin=1in,footskip=0.25in]{geometry}
\usepackage{makecell}
\usepackage[noadjust]{cite}
\usepackage{multicol}
\usepackage{graphicx}
\theoremstyle{theorem}

\newtheoremstyle{defi}
  {10pt}          
  {10pt}  
  {\rm}  
  {\parindent}     
  {\bf}  
  {. }    
  { }    
  {}     
\theoremstyle{defi}

\begin{document}

\date{}
    \maketitle
    \begin{abstract}
We explore an autonomous system analysis of dark energy models with interactions between dark energy and cold dark matter in a general systematic approach to cosmological fluids. We investigate two types of models such as local and non-local ones. In particular, a local form of interaction is directly proportional to only the energy density, while a non-local interaction is directly proportional to the energy density as well as the Hubble parameter. As a consequence, it is explicitly demonstrated that in both cases there exist the stability points in terms of cosmological parameters. This work aims at obtaining acceleration and stability using interaction models without modifying the matter or geometric component of the Universe.
    \end{abstract}

\section{Introduction}

There are large number of cosmological evidence in the form of observation which proves that our universe is presently experiencing an accelerated expansion. In the standard Einstein gravity, the late time cosmological acceleration is explained by introducing an exotic energy component with huge negative pressure, known as dark energy. In this quest, the cosmological constant $\Lambda$ was introduced as the simplest candidate of dark energy. However, it has some difficult theoretical issues such as fine tuning and cosmic coincidence problem. It is important to know the exact nature of dark energy. Since then various dynamical dark energy models have also been explored in several articles \cite{ 1,2,3,4,5,6,7,8,9,10,11,13,14,15,16,17,19,20,21,Chakraborty:2018bxh, Roy:2018eug}. Moreover to obtain late time cosmic acceleration a large scale modification of gravity has been used. At present late time cosmic acceleration is considered to be an established phenomenon but its cause is still not clearly understood. There are numerous models in Einstein gravity (with cosmological constant) and modified gravity theories which explain the phenomenon of accelerated expansion \cite{Nojiri:2010wj, DeFelice:2010aj, Capozziello:2010zz, Capozziello:2011et, Cai:2015emx,Nojiri:2017ncd, Copeland:2006wr, Bamba:2012cp, Shah, Shah1, Shah2}.

Even though $\Lambda CDM$ model is consistent with current observations, yet coincidence problem does not have a valid argument. Interaction of dark energy with dark matter is one other approach that is useful to address the mentioned problem. The interacting dark energy models have been recently proposed by several authors \cite{22,23,24,25}. The interaction between dark energy and dark matter may enhance the nature of dark matter, and also affect structure formation. The investigation of phase space analysis is the one of the most effective test for dark energy models \cite{SB, SD, Oikonomou, Chatzarakis}. If the dark energy models have $\frac{\Omega_{DE}}{\Omega_{DM}}$ of the order $1$ and an accelerated scaling attractor solution, then the coincidence problem can be resolved. The non-interacting quintessence \cite{28,29} and quintom models \cite{31,32,33} show late time accelerated attractors, and possess $\Omega_{DM}$ as 1, therefore, they do not provide an adequate solution for coincidence problem. In the literature, two forms of linear interactions have been studied namely local and non-local. Local forms of interactions are the ones which are directly proportional to energy density whereas non- local forms are directly proportional to Hubble parameter H and energy density $\rho$. In this work we consider one example of both local form which is proportional to energy density and non local form of interactions proportional to both Hubble parameter H and energy density. Many local forms interaction models have been discussed in references \cite{34, 35, 36,37}. There is also an approach to discuss the interacting term without the assumption of a specific form of interacting term which is discussed in \cite{38}. To study these interaction models we have incorporated the technique of autonomous systems. This technique gives us the qualitative behaviour of the systems without solving the equations analytically. The critical points are the zeros of the systems for the interacting dynamical model. These points can be plotted in different ways to visualize the role of interaction models. This work is organised in the following manner: In section II we establish the interacting quintessence cosmological framework and construct an autonomous dynamical system which is worthy for phase space investigation. We then discuss phase space analysis and find stationary points and their stability for interacting cosmological models. Conclusion and discussions are presented in section III.

\section{Interacting Dark Energy and Dark Matter}
Consider that our universe is well described by a flat Friedmann–Lemaitre–Robertson–Walker (FLRW) metric
\begin{equation}\label{1}
ds^2=-dt^2 + a^2(t)(dx^2+ dy^2+ dz^2)
\end{equation}
with the matter distribution obeying the perfect fluid distribution with the energy-momentum tensor
\begin{equation}\label{2}
T_{\mu\nu} = (p + \rho)u_{\mu}u_{\nu} + pg_{\mu\nu}
\end{equation}
where $u_{\mu}$ is the four velocity vector of the perfect fluid. $\rho$ and $p$ are the energy density and the thermodynamic pressure of the perfect fluid. The explicit form of the Einstein’s field equations (assuming c = 1) would be
\begin{equation}\label{3}
G_{\mu\nu} = 8\pi GT_{\mu\nu}
\end{equation}
The two independent Friedmann’s equations derived are
\begin{equation}\label{4}
H^2 = \frac{8\pi G}{3} (\rho_m + \rho_r + \rho_d)
\end{equation}
\begin{equation}\label{5}
2\dot{H} + 3H^2 = -8\pi G(p_m + p_r + p_d),
\end{equation}
where $H$ = $\dot{a}/a$ the Hubble parameter. Dot represents the derivative with respect to cosmic time t. $\rho_m$ and $\rho_d$ are the energy densities of dark matter and dark energy. $p_m$ and $p_d$ are the thermodynamic pressure of Dark Matter and Dark Energy. We assume dark matter as a pressureless dust i.e. $p_m = 0$. The dark energy satisfies the equation of state $p_d= \omega_d \rho_d$, where $\omega_d$ is the equation of state for dark energy. We consider the universe to be filled with dark matter, radiation and dark energy. We consider the interaction only between dark energy and dark matter components and write the conservation equations in the following coupled form:

\begin{equation}\label{6}
\dot{\rho}_m + 3H \rho_{m} = Q,
\end{equation}

\begin{equation}\label{7}
\dot{\rho}_d + 3H(1 + \omega_{d})\rho = -Q
\end{equation}

\begin{equation}\label{8}
\dot{\rho}_r + 4H \rho_{r} = 0,
\end{equation}
Here, $Q$ is the rate of energy density exchange between dark matter and dark energy and its positive value denotes the energy transfer from dark energy to dark matter while negative value denotes the transfer from dark matter to dark energy. $Q = 0$ would mean that there is no energy transfer between these two quantities and it would be the case similar to non interacting model.

$Q$ is assumed to be positive for the validity of the second law of thermodynamics. If we see the continuity equations \eqref{6} and \eqref{7}, the interaction between dark energy and dark matter must be a function of the energy densities multiplied by a quantity having units of the inverse of time which without the loss of generality can be chosen as the Hubble parameter. Thus interaction between dark energy and dark matter could be expressed in the most general form as $Q = Q(H, \rho_m, \rho_d)$. In various literature, the nature of Dark Energy has been studied considering different type of interactions \cite{39,40}. We consider for simplicity that the interaction is in linear combinations of the dark sector densities as \cite{34,41,42,44}.

\begin{equation}\label{9}
Q = 3\lambda_m H \rho_m + 3\lambda_d H \rho_d
\end{equation}
where $\lambda_m$ and $\lambda_d$ are dimensionless constants. As from observational point of view the interaction should be sub-dominant today \cite{24}, so, $\lambda_m$ and $\lambda_d$ are very small. The factor ‘3’ in the above expression for interaction is motivated purely from mathematical ground. This general form of interaction has been studied recently by several authors \cite{34, 41,42,44} and the particular cases $\lambda_m= \lambda_d$, in cite{46}, and $\lambda_d= 0$, in \cite{47}. Inserting \eqref{9} in the energy conservation equations \eqref{6} and \eqref{7} we have

\begin{equation}
\dot{\rho}_m + 3H \left(1- \lambda_m - \frac{\lambda_d}{u}\right)\rho_m = 0,
\end{equation}

\begin{equation}
\dot{\rho}_d + 3H \left(1 + \omega_d + \lambda_d + \lambda_m u \right)\rho_d = 0
\end{equation}
Define various density parameters as:
\begin{equation}
\Omega_{m} = \frac{\rho_{m}}{3H^2} , ~~ \Omega_{r} = \frac{\rho_{r}}{3H^2} , ~~ \Omega_{d} = \frac{\rho_{d}}{3H^2}
\end{equation}

\section{Generic autonomous system approach}
We analyze this model by considering cosmological constant as the governing force of acceleration of the universe. Hence $\omega_d = -1$. The technique of dynamical systems is used by introducing certain dimensionless parameters to study the stability of interaction model.
The dimensionless parameters introduced are:

\begin{equation}
x = \frac{\rho_{m}}{3H^2} , ~~ y = \frac{\rho_{r}}{3H^2}
\end{equation}
So we have,
\begin{equation}
\Omega_{d} = 1 - x - y
\end{equation}
We divide the above problem into three cases:

\textbf{Case 1:} $\lambda_d = 0$

When $\lambda_d = 0$, we have $Q = 3\lambda_m H \rho_m$.
Assuming this interaction term will lead us to following set of autonomous differential equations:

\begin{equation}{\label{15}}
x' = x\left( -3(1 - \lambda_m) + 3x +4y \right)
\end{equation}

\begin{equation}{\label{16}}
y' = y \left(3x+4y-4 \right)
\end{equation}

\begin{figure}[!htp]
\centering
\includegraphics[scale=0.55]{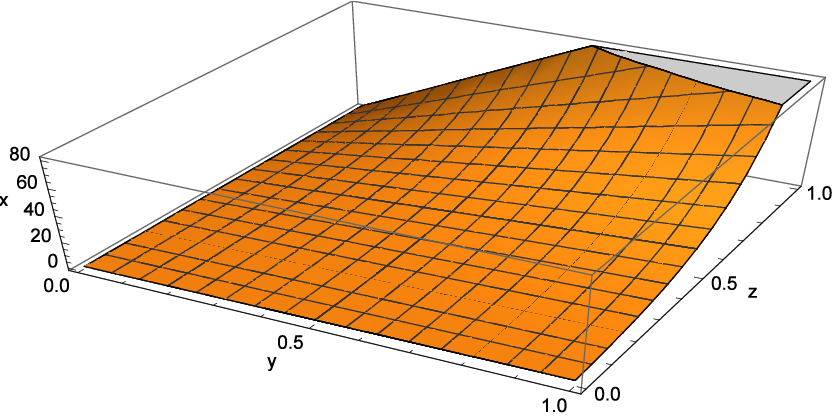}
\caption{3D plot of \eqref{b} for $\lambda_m =1$}
\label{7.8}
\end{figure}

We can solve these two equations. Let $x^{\prime}=\frac{dx}{dz}$. Then we can one times integrate Eqs.(15)-(16). As result we obtain
\begin{equation}{\label{a}}
\frac{x}{y}=e^{[C_{1}+(1+3\lambda_{m})z]}
\end{equation}
or
\begin{equation}{\label{b}}
x=ye^{[C_{1}+(1+3\lambda_{m})z]}
\end{equation}
where $C_{1}=const$. In figure \ref{7.8},  we have presented the 3-dimensional portrait of \eqref{b} in terms of coordinates $x,y,z$.

We analyze this system using dynamical system analysis. O(0,0), A(0,1) and B($1-\lambda_m$,0) are the critical points of this system. Physically (0,0) would mean that the universe is completely filled with dark energy and (0,1) would mean that universe is completely filled with radiation.

By evaluating Jacobian at the above mentioned critical points and finding its eigenvalues we get:

\begin{table}[h!]
\centering
\small\addtolength{\tabcolsep}{-5pt}
\begin{tabular}{ |c|c|c|c| }
 \hline \textbf{Point} &  \textbf{$\omega_{eff}$} & \textbf{Eigenvalues} & \textbf{Stability} \\
 \hline
 (0, 0) &  $-1$ & {-4, $-3(1-\lambda_m)$} & Stable\\
 \hline
 (0, 1) &  $\frac{1}{3}$ & {4, $1+ 3\lambda_m$} & Unstable \\
 \hline
 ($1- \lambda_m$, 0) &  $- \lambda_m$ & {$ -1 - 3\lambda_m$, $3(1-\lambda_m)$)} & Saddle Point \\
 \hline
\end{tabular}
\caption{Stability analysis for $\lambda_d =0$}
\label{7.1}
\end{table}

Phase space portrait of the dynamical system \eqref{15}-\eqref{16} is plotted now for $\lambda_m = 0.1$. Clearly origin is the stable point and ($1- \lambda_m$, 0) i.e. $(0.9,0)$ is the saddle point in this system.

\begin{figure}[!htp]
\centering
\includegraphics[scale=0.25]{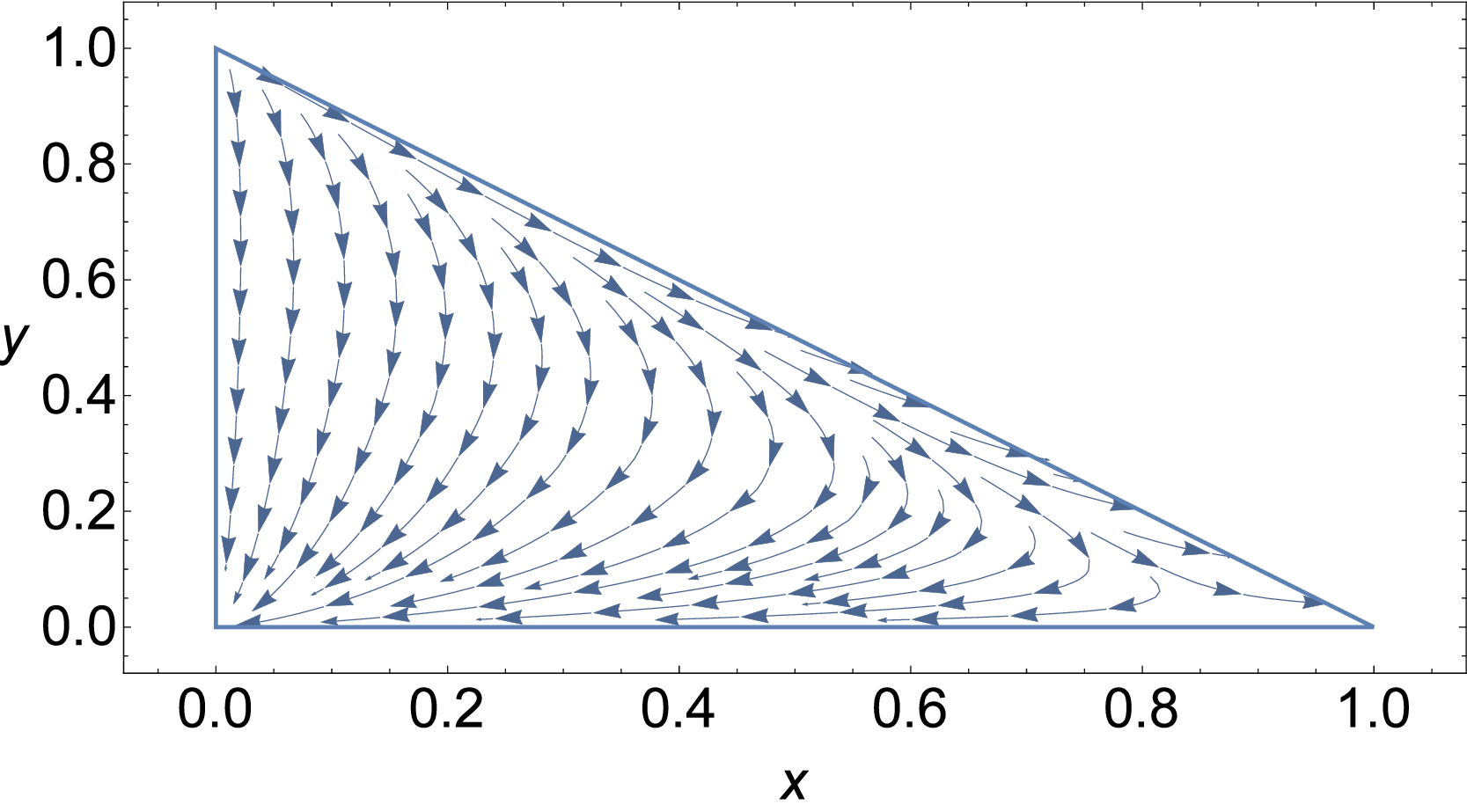}
\caption{Phase space portrait of \eqref{15} and \eqref{16}}
\label{7.1}
\end{figure}

We can also study the relative energy density of dark matter, radiation and dark energy together with effective equation of state parameter in $\Lambda CDM$ model.

\begin{figure}[!htp]
\centering
\includegraphics[scale=0.50]{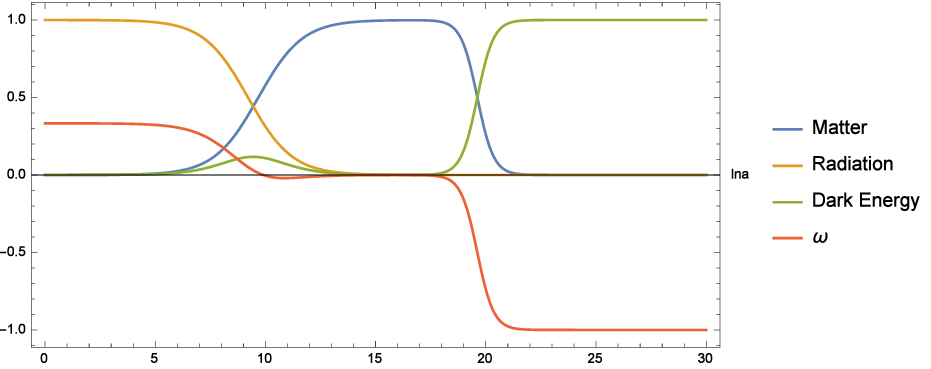}
\caption{Relative energy density for $\lambda_d = 0$}
\label{7.2}
\end{figure}

In above figure we see that during initial stage, we have universe completely filled with radiation, which then slowly reduces and matter starts to form, hence increasing its relative energy density. In the later stage dark energy starts to dominate and hence causing the accelerated expansion of the universe which we currently observe. $\omega_{eff}$ starts with $\frac{1}{3}$ as universe was completely radiation dominated then during the matter formation era, it goes down to zero and further when dark energy starts to dominates, it goes negative and ultimately reach the value of $\omega_{\Lambda}$ which is -1. \\

\textbf{Case 2:} $\lambda_d = \lambda_m = \lambda$
When $\lambda_d = \lambda_m = \lambda$, we have $Q =3 \lambda H (\rho_m + \rho_d)$. For this interaction term we arrive at following set of autonomous differential equations:

\begin{equation}{\label{17}}
x' = -3x + 3\lambda x + 3\lambda (1-x-y) + 3x^2 + 4xy
\end{equation}

\begin{equation}{\label{18}}
y' = y \left(3x+4y-4 \right)
\end{equation}
The critical points of this system are \textbf{A} $(0,1)$, \textbf{B} $\left( \frac{1}{2} \left( 1 - \sqrt{1 - 4\lambda} \right), 0 \right)$ and \textbf{C} $\left( \frac{1}{2} \left( 1 + \sqrt{1 - 4\lambda} \right), 0 \right)$. We here note that the critical points are real only when $\lambda \leq 1/4$. Eigenvalues of point \textbf{A} are $\frac{1}{2} \left( 5 - 3 \sqrt{1 - 4\lambda} \right)$ and $\frac{1}{2} \left( 5 + 3 \sqrt{1 - 4\lambda} \right)$, eigenvalues of point \textbf{B} are $-4 + \frac{3}{2} \left( 1 - \sqrt{1 - 4\lambda} \right)$ and $-3 + 3 \left( 1 - \sqrt{1 - 4\lambda} \right) $ and eigenvalues of point \textbf{C} are $-4 + \frac{3}{2} \left( 1 + \sqrt{1 - 4\lambda} \right)$ and $-3 + 3 \left( 1 + \sqrt{1 - 4\lambda} \right) $. The brief analysis of these critical points is as follows:
\begin{table}[h!]
\centering
\small
\hskip-1.0cm \begin{tabular}{ |c|c|c| }
 \hline \textbf{Point} &  \textbf{$\omega_{eff}$} & \textbf{Stability} \\
 \hline
 \textbf{A} & $\frac{1}{3}$ & Unstable\\
 \hline
\textbf{B} &  $-\frac{1}{2} \left( 1 + \sqrt{1 - 4\lambda} \right)$ & Stable \\
 \hline
\textbf{C} &  $-\frac{1}{2} \left( 1 - \sqrt{1 - 4\lambda} \right) $ & Saddle Point \\
 \hline
\end{tabular}
\caption{Stability analysis for $\lambda_d = \lambda_m$}
\label{7.2}
\end{table}

Phase space portrait of the dynamical system \eqref{17}-\eqref{18} is plotted now for the value $\lambda = 0.1$. Clearly point \textbf{B} i.e. $(0.1127,0)$ is the stable point in this system and point \textbf{C} i.e. $(0.8873,0)$ would be saddle point of this system.

\begin{figure}[!htp]
\centering
\includegraphics[scale=0.25]{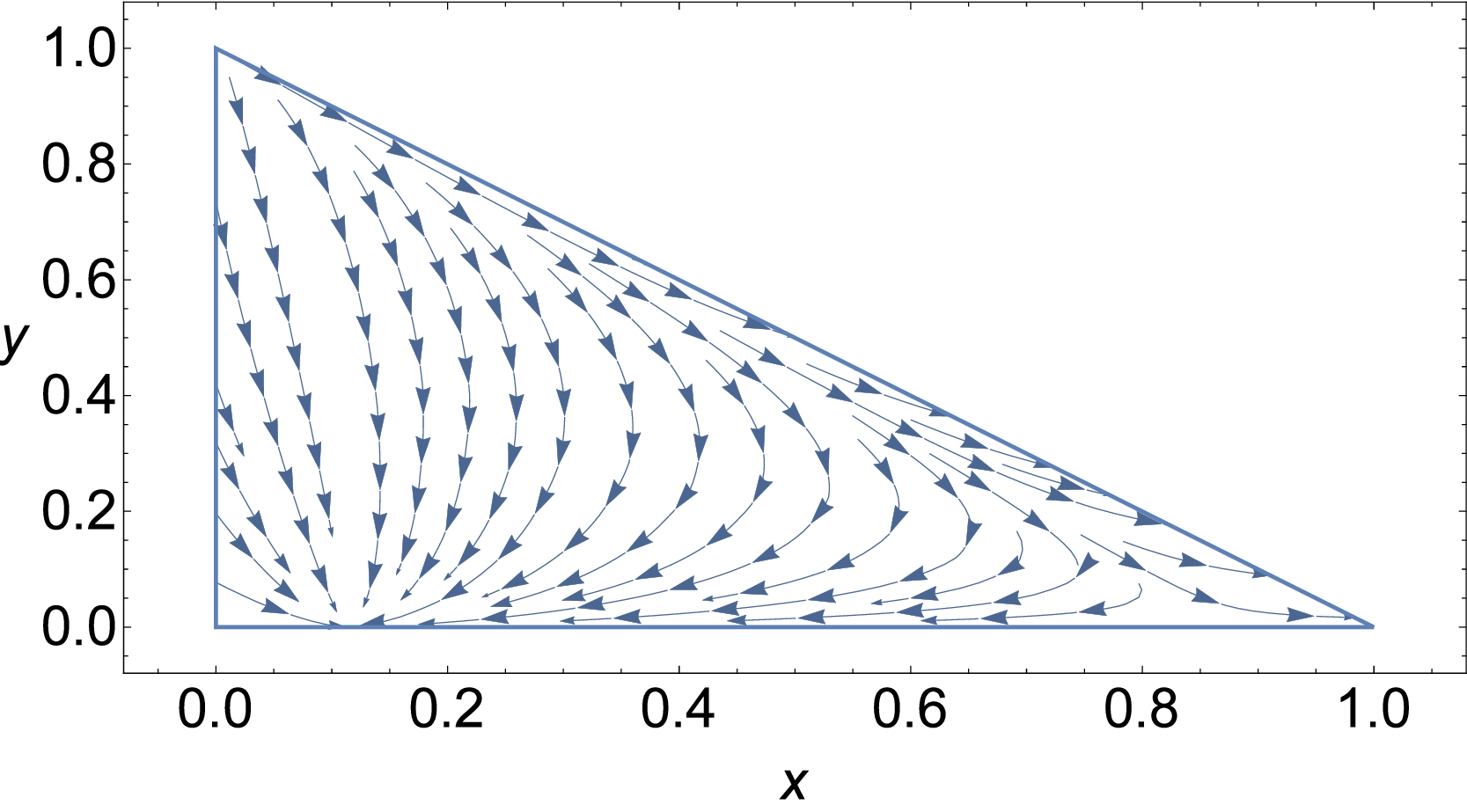}
\caption{Phase space portrait of \eqref{17} and \eqref{18}}
\label{7.3}
\end{figure}

\textbf{Case 3:} $\lambda_m = 0$
When $\lambda_m = 0$, we have $Q =3 \lambda_d H\rho_d $. For this interaction term we arrive at following set of autonomous differential equations:

\begin{equation}{\label{19}}
x' = -3x + 3\lambda_d (1-x-y) + 3x^2 + 4xy
\end{equation}

\begin{equation}{\label{20}}
y' = y \left(3x+4y-4 \right)
\end{equation}
The critical points of this system are \textbf{O} $(\lambda_d, 0)$, \textbf{A} $ (0,1)$ and \textbf{B} $(1,0)$. The brief analysis of these critical points is as follows:

\begin{table}[h!]
\centering
\small\addtolength{\tabcolsep}{-5pt}
\begin{tabular}{ |c|c|c|c| }
 \hline \textbf{Point} &  \textbf{$\omega_{eff}$} & \textbf{Eigenvalues} & \textbf{Stability} \\
 \hline
 ($\lambda_d$, 0) &  $-1 + \lambda_d$ & $3 \lambda_d -4, 3 \lambda_d -3$ & Stable\\
 \hline
 (0, 1) &  $\frac{1}{3}$ & {1, 4 - 3$\lambda_d$} & Unstable \\
 \hline
 (1, 0) &  $0$ & {-1, 3 - 3$\lambda_d$} & Saddle Point \\
 \hline
\end{tabular}
\caption{Stability analysis for $\lambda_m = 0$}
\label{7.3}
\end{table}

Phase space portrait of the dynamical system \eqref{19}-\eqref{20} is plotted now for the value $\lambda = 0.1$. Clearly point \textbf{O}: (0.1,0) is the stable point in this system and point \textbf{B}: (1,0) would be saddle point of this system.

\begin{figure}[!htp]
\centering
\includegraphics[scale=0.25]{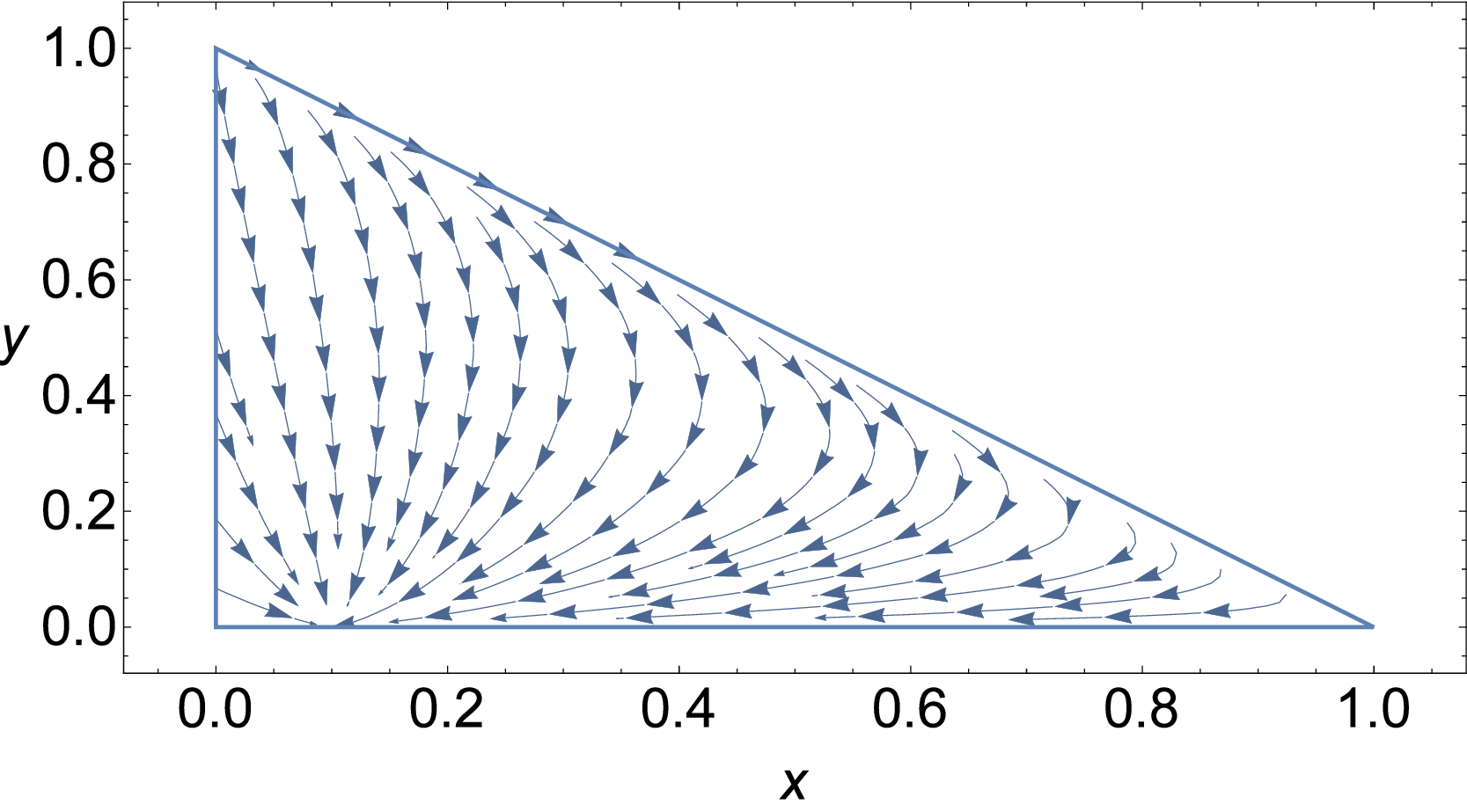}
\caption{Phase space portrait of \eqref{19} and \eqref{20}}
\label{7.4}
\end{figure}

As the next form of interaction let us consider $Q = \alpha \dot{\rho}_m$. This interaction term will lead to the following set of dynamical system equations.

\begin{equation}{\label{21}}
x' = \frac{-3x}{1-\alpha} + 3x^2 + 4xy
\end{equation}

\begin{equation}{\label{22}}
y' = y \left(3x+4y-4 \right)
\end{equation}

\begin{figure}[!htp]
\centering
\includegraphics[scale=0.55]{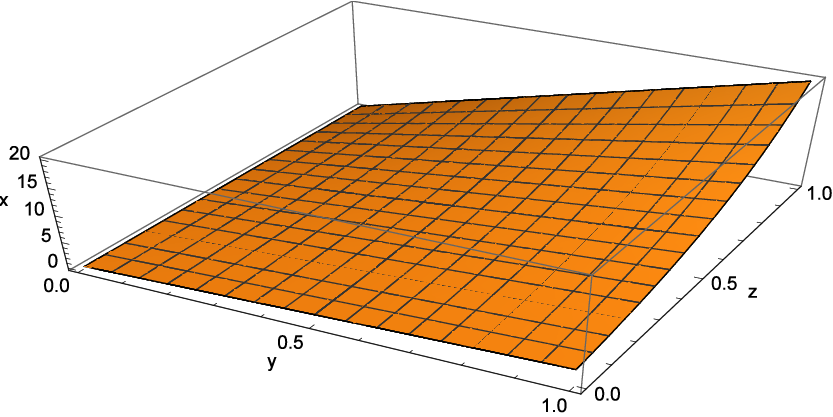}
\caption{3D plot of \eqref{d} for $\alpha=0.5$}
\label{7.9}
\end{figure}

We can solve these two equations. Let $x^{\prime}=\frac{dx}{dz}$. Then we can one times integrate Eqs.(21)-(22). As result we obtain
\begin{equation}{\label{c}}
\frac{x}{y}=e^{[C_{2}-(\frac{1-4\alpha}{1-\alpha})z]}
\end{equation}
or
\begin{equation}{\label{d}}
x=ye^{[C_{2}-(\frac{1-4\alpha}{1-\alpha})z]}
\end{equation}
where $C_{2}=const$. In Figure \ref{7.9},  we have presented the 3-dimensional portrait of \eqref{d} in terms of coordinates $x, y, z$.

We analyze this system using dynamical system analysis. The critical point of the system are O: $(0, 0)$, A: $(0, 1)$ and B: $(\frac{1}{\alpha -1}, 0)$. The brief analysis of these critical points is as follows:

\begin{table}[h!]
\centering
\small\addtolength{\tabcolsep}{-5pt}
\begin{tabular}{ |c|c|c|c| }
 \hline \textbf{Point} &  \textbf{$\omega_{eff}$} & \textbf{Eigenvalues} & \textbf{Stability} \\
 \hline
 $(0, 0)$ &  $-1$ & $-4, \frac{3}{\alpha -1}$ & Stable for $\alpha < 1$\\
 \hline
 $(0, 1)$ &  $\frac{1}{3}$ & $4, \frac{4\alpha - 1}{\alpha -1}$ & Unstable \\
 \hline
 $(\frac{1}{\alpha -1}, 0)$ &  $\frac{2 - \alpha}{\alpha - 1}$ & $\frac{3}{\alpha -1}, \frac{1-4\alpha}{\alpha -1}$ & Stable for $\alpha > 1$ \\
 \hline
\end{tabular}
\caption{Stability analysis for $Q = \alpha \dot{\rho}_m$}
\label{7.4}
\end{table}

Phase space portrait of the dynamical system \eqref{21}-\eqref{22} is plotted now for the value $\alpha = 0.5$. Clearly point \textbf{O}: (0,0) is the stable point in this system and point \textbf{A}: (0,1) would be unstable point of this system.
\begin{figure}[!htp]
\centering
\includegraphics[scale=0.25]{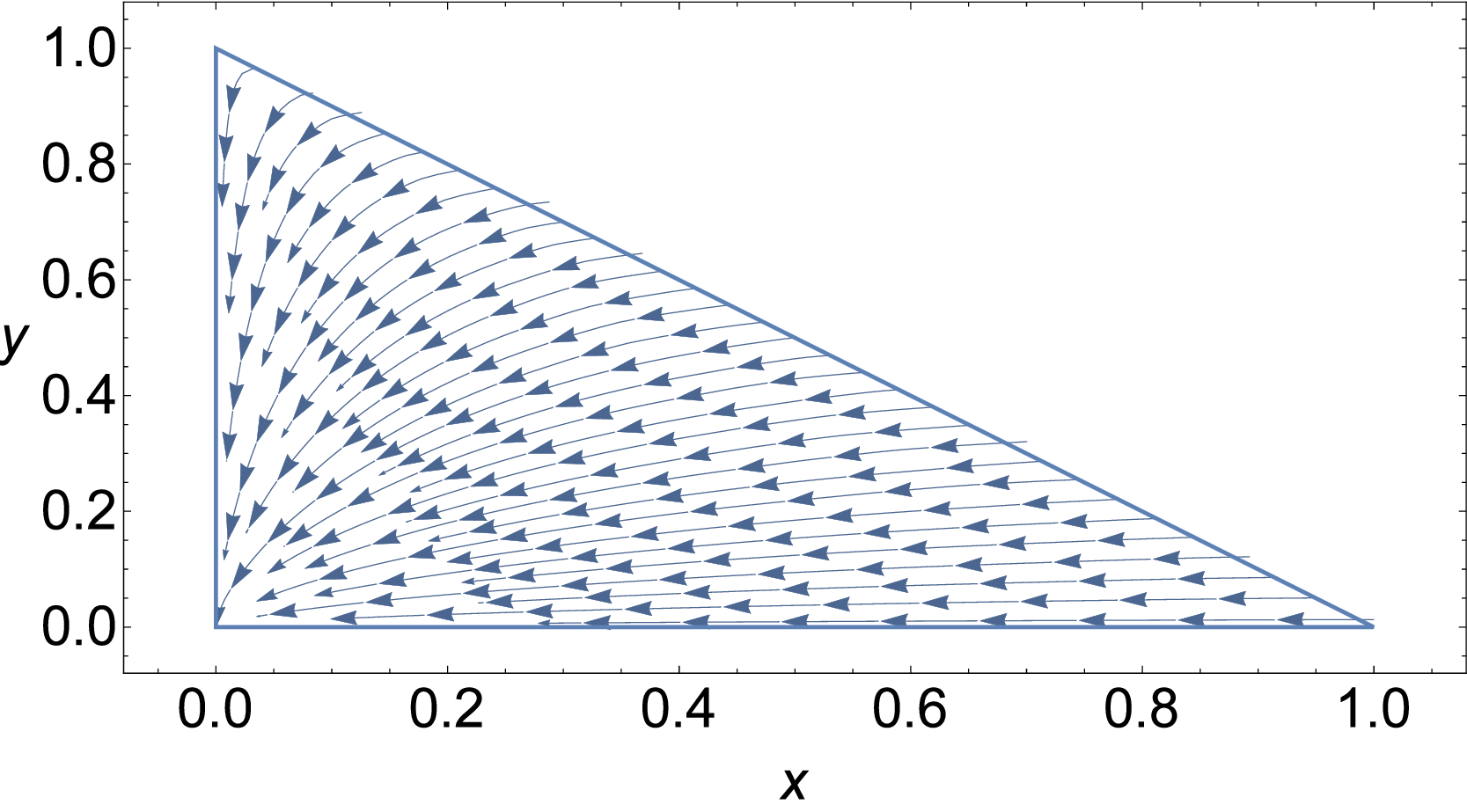}
\caption{Phase space portrait of \eqref{21} and \eqref{22} for $\alpha = 0.5$}
\label{7.5}
\end{figure}

We can also study the relative energy density of dark matter, radiation and dark energy together with effective equation of state parameter in $\Lambda CDM$ model.

\begin{figure}[!htp]
\centering
\includegraphics[scale=0.5]{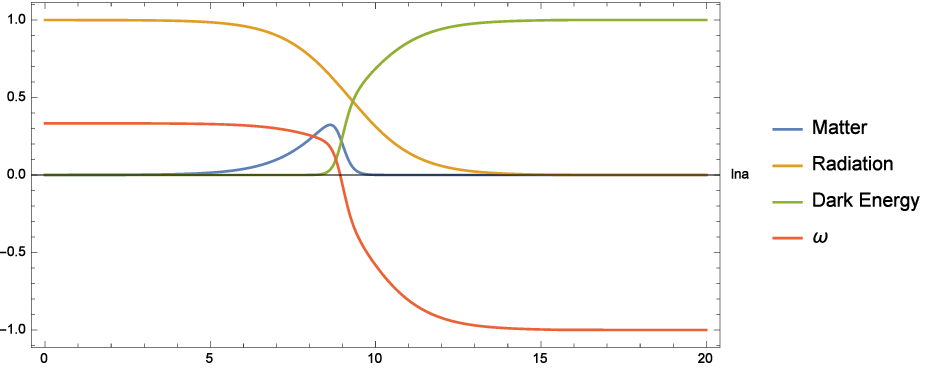}
\caption{Relative energy density for $Q = \alpha \dot{\rho}_m$}
\label{7.6}
\end{figure}

In above figure we see that during initial stage, we have universe completely filled with radiation, which then slowly reduces and matter starts to form, hence increasing its relative energy density. However matter starts to diminish soon after that and dark energy starts to dominate and hence causing the accelerated expansion of the universe which we currently observe. $\omega_{eff}$ starts with $\frac{1}{3}$ as universe was completely radiation dominated then during the matter formation era, it reduced and further when dark energy starts to dominates, it goes negative and ultimately reach the value of $\omega_{\Lambda}$ which is -1. \\

Phase space portrait of the dynamical system \eqref{21}-\eqref{22} is plotted now for the value $\alpha = 2$. Clearly point \textbf{B}: (1,0) is the stable point in this system and point \textbf{A}: (0,1) would be unstable point of this system.
\begin{figure}[!htp]
\centering
\includegraphics[scale=0.25]{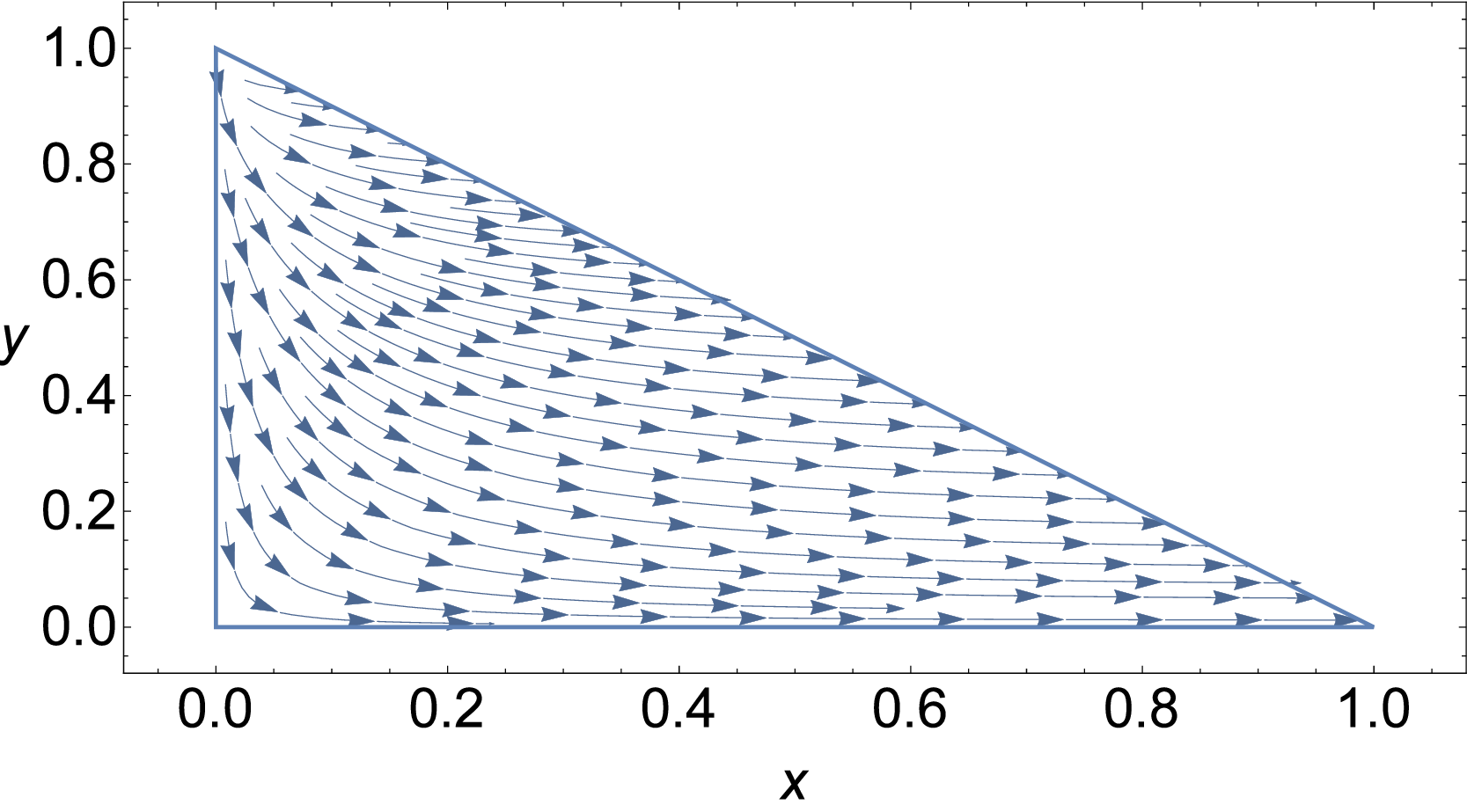}
\caption{Phase space portrait of \eqref{21} and \eqref{22} for $\alpha = 2$}
\label{7.7}
\end{figure}
Hence as obtained from the mathematical expression point O is stable when $\alpha < 1$ and point B is stable when $\alpha > 1$. Whereas point A is always unstable.

\section{Discussion and Conclusion}
In this work we discuss the interaction between dark matter and dark energy in the general theory of gravity using the approach of dynamical system of analysis. The universe is assumed to be described by the flat FLRW metric. The coupled form of the conservation equations are mentioned with $Q$ as the interaction term. Interaction term indicates the rate of energy transfer between dark matter and dark energy. Positive $Q$ indicates that the energy is transferred from dark energy to dark matter and negative $Q$ indicates that the energy is transferred from dark matter to dark energy. $Q=0$ indicates the model similar to the no-interaction model. We study this with autonomous system analysis which helps us to obtain cosmological parameters for the interaction model.

We have discussed two types of models namely local and non-local. An interaction is said to be of local form if it is directly proportional to energy density only and a non-local interaction is directly proportional to energy density as well as Hubble parameter $H$. The non-local interaction which is studied in this work is $Q = 3\lambda_m H \rho_m + 3\lambda_d H \rho_d$. This is one of the most general interaction model studied in the literature. Further for this interaction we have studied three different cases. Case 1 discusses $\lambda_d = 0$, $\lambda_d = \lambda_m$ is studied in case 2 and case 3 discusses $\lambda_M = 0$. In each of these cases, we first obtain the set of autonomous differential equations. From this, we get the critical points of the system. The eigenvalues of these critical points gives a conclusive evidence about the stability of the critical point and $\omega_{eff}$ is also evaluated in each of the case to study the acceleration at that critical point. We also obtain a 2D phase space portrait to show the stability. The $x$ coordinate of the critical point describes the energy density of matter part of the universe. Similarly $y$ coordinate describes the energy density of radiation part of the universe. It must be noted that the point $(0,0)$ describes the universe completely dominated by dark energy. We also study the relative energy density of the universe at various times. For a non-local interaction form, we obtain a universe which begins with complete radiation dominance and later when the matter formation begins the radiation part decreases. At one stage the universe is completely matter dominated and other components are zero. At the end stage, the universe is completely dark energy dominated and other two components are zero which leads to the late time acceleration of the universe. We note here that although the completely radiation dominated universe and completely dark energy dominated match with the observational evidence however a completely matter dominated universe does not support any observational evidence. Later the local interaction form $Q =\alpha \dot{\rho}_m$. is discussed. All the calculations done for local form of interaction are also done for the non-local form of interaction. Here different stable points are achieved for different values of $\alpha$. In the non-local form of interaction when we evaluate the relative energy density, we observe that beginning of the universe is with the radiation dominated universe and at the end it is completely dark energy dominated universe which is similar to what was observed in local form also. But here we saw that universe is never completely dominated by the matter part. This plot makes this form of interaction very similar to the observational evidence.

From all the calculations and discussions, we could conclude that for the linear interaction models acceleration and stability phase could be observed  without modifying the geometry part of the Einstein's field equations. In this work, we obtain stable points which corresponds to a particular physical state of the Universe and these points are the attractors for our model. Subsequently, we observed that a physical system with some initial conditions, will always try to attend the stable state which is evident from the plots (Fig. 2, 4, 5, 7 and 9).

In future, other interactions models could be studied. There are several non-linear form of interactions which might also help in theoretical studies of the cosmological models. Moreover, these interactions could also be studied in the modified gravity which can help in further understanding of local, non-local, linear and non-linear interactions. In particular, there are studies on non-local gravity under Integal Kernel Theories of Gravity (IKGs) which take into account integral kernels of differential operators, such as the inverse d'Alembert operator \cite{50} and \cite{51}. This approach of IKGs emerge in searching for unitary and renormalizability of quantum gravity models. Non-local gravity models are being considered as a natural mechanism to address dark energy dynamics and dark matter issue in the large scale structure \cite{52} and \cite{53}. In future a comparative study can be done with various interaction models and non-local gravity. We also plan to investigate the method of dominant balances to see whether such cosmic singularities can occur in this framework like \cite{48} and \cite{49}. 

\section*{Acknowledgments}
The work of PS is supported by Department of Science and Technology (DST), Govt. of India via INSPIRE Fellowship (Ref. No. IF160358) and VIT AP University. The work of KB has partially been supported by the JSPS KAKENHI Grant Number JP21K03547. The work of RM is supported by the Grant AP09261147 of the MES Of Kazakhstan. We acknowledge the reviewers for giving their valuable comments and suggestions to improve the manuscript.

\end{document}